# Bottom-loading dilution refrigerator with ultra-high vacuum deposition capability

L. M. Hernandez and A. M. Goldman

School of Physics and Astronomy, University of Minnesota

Minneapolis MN, 55455

A Kelvinox 400 dilution refrigerator with the ability to load samples onto the mixing chamber from the bottom of the cryostat has been combined with an ultrahigh-vacuum (UHV) deposition chamber equipped with molecular beam sources. The liquid helium cooled sample transfer mechanism is used in a manner that allows films to be grown on substrates which are kept at temperatures of order 8K with chamber pressures in the $10^{-9}$ - $10^{-10}$ Torr range. This system facilitates the growth of quench-condensed ultrathin films which must always be kept below ~ 12K in a UHV environment during and after growth. Measurements can be made on the films down to millikelvin temperatures and in magnetic fields up to 15 T.

INTRODUCTION

Disordered ultra-thin films, i.e., quench-condensed films, are usually produced by deposition onto cooled substrates. In order to obtain reliable data with such films, one must employ precise deposition techniques, carefully controlling deposition conditions, and precisely monitoring film history. Deposition in a vacuum environment, which is not clean, can introduce substantial contamination that will result in substantial, but unknown variations of structure and composition. In addition, if such films, during deposition, are raised above a specific temperature that is material dependent, they anneal, and their microstructure and hence the physics which governs their behavior will change. For measurements of such films to be carried out at very low temperatures, one must combine dilution refrigeration and UHV techniques. Using ideas that are both a simplification and an elaboration of an earlier apparatus,[1] we have constructed a system with the ability to grow ultrathin films in a highly controlled manner, on substrates held below 12K. The substrates can then be transferred to a Kelvinox 400 $^3$He-$^4$He dilution refrigerator[2] for measurements in the millikelvin regime, while constantly held at low temperature and in a UHV environment. There is a long history of research on quench-condensed films beginning with the work of Shal'nikov.[3-9] In a direct sense the system described here represents a refinement of what was done earlier.

I.  DESIGN

A schematic of the apparatus is shown in Fig. 1. The ultrahigh vacuum (UHV) space is pumped by a 170-l/s turbomolecular pump and a 220l/s differential ion pump. The sample itself is mounted on a conical shaped copper slug provided with slip ring electrical contacts following a standard design of Oxford Instruments.[2] This slug is attached to a liquid helium cooled transfer stick. Transfer to the low temperature environment is accomplished by screwing the slug into a cold finger, which extends from the bottom of the mixing chamber of the refrigerator. This process also unscrews the transfer stick from the slug, leaving it attached to the mixing chamber. The process can be reversed. Screwing the transfer stick into the slug does this. This in turn detaches it from the cold finger. The sample, when translated vertically downward a distance of approximately 1 meter, is then positioned in the deposition chamber. Once the deposition process is completed the sequence can be reversed and the film, again attached to the mixing chamber, can be measured. There are a series of doors attached at 77K, 4K, and at the temperature of the heat exchangers in the refrigerator. These are pushed open during the movement of the transfer stick. They are closed when the sample is attached to the refrigerator, and the transfer stick is lowered below a valve that effectively isolates the vacuum chamber of the refrigerator from the growth chamber and sample transfer assembly.

With the dewar cooled to 4.2K, the base pressure of the growth chamber was $2 \times 10^{-9}$ Torr. Mass spectrometer analysis showed a dominant $H_2$ peak, but also a weak series of peaks associated with air. The origin of these air peaks is a "virtual" leak stemming from the shaft in the liquid Helium cooled siphon. In order to facilitate in-situ rotation of the sample on the slug, a long flexible metal wire runs down a shaft in the center of the siphon. This is sealed by a series of Viton O-rings at the base of the siphon. The conductance through this tube is 0.0125l/s. However, as will be discussed later, this "virtual" leak does not compromise the vacuum integrity during film growth or measurement.

The growth chamber itself was designed for ease of operation and flexibility. Two custom made chambers, connected via a hydraulically formed bellows, form the main unit. The larger chamber contains two MBE furnaces, which are individually shrouded by liquid nitrogen jackets in order to reduce outgassing during deposition. The furnaces themselves are commercial

MBE effusion cells that are used with pyrolitic boron nitride crucibles. The furnaces are powered by a DC supply, which is controlled by a proportional-integral-differential (PID) feedback loop. This allows the deposition process to be carried out at with a high level of stability at rates as low as 0.01A/sec. An eight-inch gate valve is situated between this chamber and the sample transfer chamber in order to allow repair work, or changing of evaporation sources without disturbing the vacuum in the refrigerator. A small chamber is mounted directly below the refrigerator entry port. The sample is positioned in this chamber during film growth. It is provided with a liquid nitrogen cooled shroud to minimize the heat load on the helium-cooled transfer stick. There is a small opening in the shroud that allows the vapor stream to impinge on the substrate. A residual gas analyzer monitors vacuum quality, and a calibrated, quartz crystal oscillator monitors the evaporation rate.

In addition, a load lock is mounted below this chamber. This permits one to change samples and begin a new experimental run without having to warm the refrigerator. A number of spare ports also permit optical access to the sample while it is attached to the transfer siphon and is at a temperature of ~8K. These can be used for optical excitation studies that might require a sample to remain cold and under UHV conditions.

II. OPERATION

The various vacuum chambers are pumped initially using the turbomolecular pump. Once the pressure is sufficiently low, the 220 l/s differential ion pump is turned on and the turbomolecular pump is valved off. The system is then baked at temperatures no higher than 150 $^0$C to avoid melting any Viton O-rings. The evaporation sources, which are conventional commercial Knudsen cells, are also outgassed at operating temperature during bakeout. Once the pressure stops changing, the dewar is cooled to 4.2K, which eliminates outgassing from the refrigerator insert which cannot be baked. Baking is continued for another 24 hours.

To begin the deposition procedure, the shroud surrounding the furnace is filled with liquid nitrogen and the furnace is slowly ramped to its operating temperature. A second liquid nitrogen shroud, which prevents radiative heating of the sample, is also filled. The sample slug with the substrate is then attached to the refrigerator. The $He^3$-$He^4$ mixture is removed from the refrigerator to prevent any flash boiling when the sample transfer siphon is attached to the slug affixed to the mixing chamber. Once the shrouds have cooled and the furnace is at its operating temperature, the sample transfer siphon is cooled with liquid helium. After approximately 30 minutes the thermometer mounted in the siphon reaches its base temperature of slightly more than 4.2K. At this point the virtual leak which originates from the shaft of the siphon is effectively shut off by the cryopumping by the helium reservoir in the siphon. Residual gas analysis typically shows a $H_2$ peak in the low $10^{-9}$ Torr range (due to outgassing from the hot furnaces).

At this point the sample transfer siphon is moved up to the base of the sample slug which is threaded. The welded stainless steel bellows allows vertical motion of about one meter. In order to screw the slug onto the siphon, and remove it from the mixing chamber, ten rotations must be carried out. A hollow-shaft ferrofluidic feedthrough from Advanced Fluid Systems[10] model CF63-H7/8-SHCW) allows for the continuous rotation while maintaining UHV conditions. The carrier fluid for the magnetic particles in the feedthrough is a synthetic hydrocarbon, which has a vapor pressure of $3\times10^{-11}$ Torr. Due to the possibility of it being in a high field environment, it was shielded, and the AlNiCo magnets were replaced with NdFeBr magnets. It should be mentioned that every time the slug is screwed onto the refrigerator (or the siphon) that a considerable amount of frictional heating is observed. The temperature as monitored on the mixing chamber rises to 8K when the slug is screwed onto the refrigerator and the siphon is removed. It is therefore essential to sustain the cooling of the siphon during rotation in order to avoid irreversible annealing the film.

Once the slug is affixed to the siphon it is moved vertically downward and position inside a keyway mounted on the 4.2K tail of the dewar. The sample slug is then tightened onto the siphon with a calibrated torque wrench to 7 N-m, in order to prevent a loss of torque upon successive loading and unloading. The slug is then moved the rest of the way down to the deposition chamber. Material is deposited, and the slug with the substrate attached, is returned to the refrigerator for measurements.

The sample is cooled using a Kelvinox-400 dilution refrigerator supplied by Oxford Instruments. The base temperature of the refrigerator is 4 mK as measured with a nuclear orientation thermometer mounted on the mixing chamber. Electron temperatures have been measured on the slug down to 50 mK with a calibrated $RuO_x$ resistance thermometer. Temperatures the order of 10 mK have been achieve, with the precise limit being uncertain because of questions relating to thermometer calibration.

All electrical lines to the sample were filtered using $\pi$–section filters with a cutoff frequency of about 10kHz, soldered into a Cu-ground plane. In addition 10k$\Omega$ metal film resistors were placed on either side of the ground plane, which decreased the cutoff frequency to about 500Hz. In studies carried out thus far, a filtered Keithley 220 DC current source and a Keithley 182 nanovoltmeter were used to make low noise, four probe measurements. The Keithley 220 current source was filtered with an effective cutoff frequency of about 10Hz. The measuring system was also optically isolated from the controlling computer.

The dilution refrigerator is also equipped with a superconducting solenoid that can provide magnetic fields up to 15T and the sample rotation capability of the transfer stick and sample slug permit the angle of the film with respect to the field direction to be changed.

III. DISCUSSION

The study of quench-condensed ultrathin films requires the ability to deposit films in a well-controlled manner while maintaining a clean vacuum environment to prevent contaminants from influencing the composition and structure of the film. We have successfully constructed an apparatus, which satisfies these conditions. The apparatus has the further ability to grow films on substrates held near liquid helium temperatures and then measure them at dilution refrigerator temperatures and in very large magnetic fields.

Up to now, we have grown films of a-Bi on a 6Å thick amorphous Ge underlayer in order to study the evolution of superconductivity from the initial insulating state of these films. With this apparatus it will be possible to study transport properties of ultrathin films down to lower temperatures than have been previously achieved in experiments with similar films. With modification of the sample slug, it may be possible to carry out film growth at elevated temperatures in an ultrahigh vacuum environment. This would permit epitaxial films to be prepared and studied *in situ.* The sample insertion system would also lend itself to connection with other UHV systems so that a series of measurements could be carried out on a single film without removing it from vacuum. Finally, the refrigerator and magnet assembly is quite rigid because of the bottom loading design. The cryostat is fastened both at the top of the dewar in a conventional manner and at the bottom with various bellows assemblies. The dewar structure itself is actually suspended from an air table and the pumps are located in a remote room. As a consequence the vibration level may be low enough to permit the use of a miniature scanning tunneling (STM) microscope built into one of the sample slugs. Such a configuration could be used for STM studies at dilution refrigerator temperatures at substantial magnetic fields.

ACKNOWLEDGEMENTS

We would like to thank the machine shop of the School of Physics and Astronomy for their outstanding work on this system. We would also like to thank Drs. W. Huber, A Bhattacharya, and C. Christiansen for numerous helpful discussions. This work supported by the National Science Foundation under Grants NSF/DMR-9876816 and NSF/DMR-9412584.

FIGURE CAPTIONS

Fig. 1. Schematic of apparatus: (A) Mixing Chamber; (B) Superconducting Magnet; (C) Sample Slug; (D) Radiation Doors; (E) Liquid Nitrogen Cooled Shroud; (F) Ferrofluidic Feedthrough; (G) Liquid Helium Cooled Transfer Siphon; (H) Sample Slug Keyway; (I) Optical Access; (J) Liquid Nitrogen Cooled Jacket; (K) Quartz Crystal Thickness Monitor; (L) K-cell furnace; (M) Ion Pump

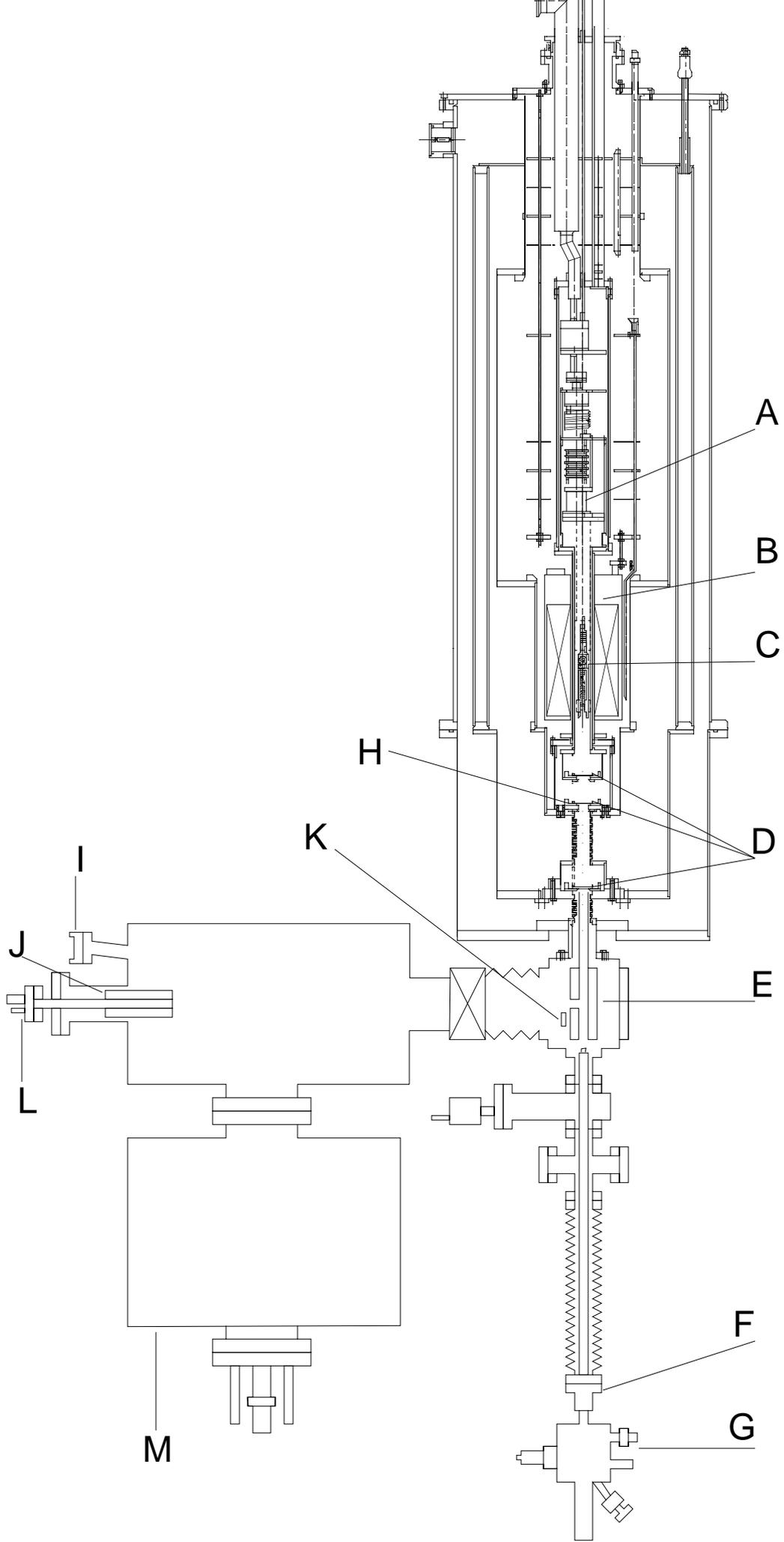